# Three-dimensional coherent X-ray diffraction imaging of a whole, frozen-hydrated cell

Jose A. Rodriguez[1], Rui Xu[2], Chien-Chun Chen[2], Zhifeng Huang[2], Huaidong Jiang[3], Kevin S. Raines[2†], Daewoong Nam[4], Allan L. Chen[5], Alan P. J. Pryor[2], Lutz Wiegart[6§], Changyong Song[4], Anders Madsen[6‡], Yuriy Chushkin[6], Federico Zontone[6], Peter J. Bradley[5], Jianwei Miao[2*]

[1]*Department of Biological Chemistry, UCLA-DOE Institute for Genomics and Proteomics, University of California, Los Angeles, California, 90095, USA.* [2]*Department of Physics & Astronomy and California NanoSystems Institute, University of California, Los Angeles, California, 90095, USA.* [3]*State Key Laboratory of Crystal Materials, Shandong University, Jinan 250100, China.* [4]*RIKEN SPring-8 Center, Kouto 1-1-1, Sayo, Hyogo 679-5148, Japan.* [5]*Department of Microbiology, Immunology, and Molecular Genetics, University of California, Los Angeles, California, 90095, USA.* [6]*European Synchrotron Radiation Facility (ESRF), 6 rue Jules Horowitz, BP 220, 38043 Grenoble Cedex 9, France.* [†]*Present address: Department of Applied Physics, Stanford University, Stanford, CA 94305.* [§]*Photon Sciences Directorate Brookhaven National Laboratory, Upton, New York 11973, USA.* [‡]*Present address: European X-Ray Free Electron Laser, Albert-Einstein-Ring 19, 22761 Hamburg, Germany.*

*Correspondence and requests for materials should be addressed to J.M. (e-mail: miao@physics.ucla.edu).*

**A structural understanding of whole cells in three dimensions at high spatial resolution remains a significant challenge and, in the case of X-rays, has been limited by radiation damage. By alleviating this limitation, cryogenic coherent diffraction imaging (cryo-CDI) could bridge the important resolution gap between optical and electron**

**microscopy in bio-imaging. Here, we report for the first time 3D cryo-CDI of a whole, frozen-hydrated cell – in this case a *Neospora caninum* tachyzoite – using 8 keV X-rays. Our 3D reconstruction reveals the surface and internal morphology of the cell, including its complex, polarized sub-cellular architecture with a 3D resolution of ~75-100 nm, which is presently limited by the coherent X-ray flux and detector size. Given the imminent improvement in the coherent X-ray flux at the facilities worldwide, our work forecasts the possibility of routine 3D imaging of frozen-hydrated cells with spatial resolutions in the tens of nanometres.**

Microscopy has transformed our understanding of biology and medicine. By using novel imaging technologies and labelling techniques, optical microscopy can routinely study dynamic processes in living cells[1]. The typical resolution for optical microscopy is around 200 nm, although better resolution can be achieved in certain cases by using super-resolution optical approaches[2]. To achieve considerably higher resolution, electron microscopy is the method of choice. Electron tomography is currently the highest resolution imaging technique available to study non-identical structures in three dimensions[3-5]. For radiation-hard inorganic materials, atomic scale 3D resolution has recently been achieved[3,4]. For biological specimens such as pleomorphic macromolecular assemblies, viruses, organelles and cells, the resolution is currently limited to 3–5 nm by radiation damage to the specimens[5]. However, the main limitation of electron microscopy is its applicability to thin or sectioned samples (typically $\leq$ 0.5 μm) due to the relatively small penetration depth of electrons[3-5]. There is thus an important gap between optical and electron microscopy in terms of spatial resolution, sample thickness, labelling requirement, contrast mechanism and quantitative capability. Coherent diffraction imaging (CDI) is a very promising imaging modality to bridge this gap for the following reasons[6,7]: First, due to the large penetration depth of X-rays, CDI can be used to image whole biological cells and biomaterials without the requirement of sectioning[8-24].

Second, compared to super-resolution optical microscopy[2], which allows for the selection of the fluorescently labelled molecules or molecular assemblies, CDI is based on the intrinsic density variations of biological specimens and accordingly enables quantitative 3D imaging of the entire contents of cells, cellular organelles and biomaterials in their natural contrast[9-16,18-24]. Third, compared to zone-plate X-ray microscopy[25-27], CDI avoids the use of X-ray lenses and its resolution is only limited by the radiation damage imparted on biological specimens.

Since its first experimental demonstration in 1999[6], CDI has been employed to image a variety of biological specimens including biomaterials, whole cells, cellular organelles and viruses, utilizing synchrotron radiation[8-19] and X-ray free electron lasers[20-24]. However, radiation damage has limited the applicability of CDI to high-resolution 3D imaging of biological samples[28,29]. One solution to alleviate the radiation damage problem is to keep the samples at cryogenic (liquid nitrogen) temperatures. Previous studies have suggested that cryo-CDI may be applied to imaging frozen-hydrated cells with a 3D resolution of ~5-10 nm[28,29]. While cryo-CDI has been demonstrated in two dimensions[13,14], it has thus far defied any experimental attempts to be achieved in three dimensions. Here, we report the first 3D cryo-CDI of a whole, frozen-hydrated cell at ~75-100 nm resolution. We chose to image the protozoan parasite *Neospora caninum*, which infects a wide variety of mammals and causes abortion in cattle and neuromuscular disease in dogs[30]. *N. caninum* is a relative of the human pathogen *Toxoplasma gondii*, which causes disease in immunocompromised patients and neonates, and also a relative of *Plasmodium falciparum*, the causative agent of malaria[30]. A typical *N. caninum* cell has a wedge-shaped body approximately 1-3 μm wide and 4-6 μm long. Like other apicomplexans, *N. caninum* undergoes several stages of development during its normal lifecycle. For the purpose of this study, we targeted imaging of the tachyzoite, a relevant stage in the parasite lifecycle during which it replicates quickly and infects most cell types within its host[31]. In spite of its small size, the *N. caninum* parasite hosts a remarkable array of subcellular compartments including rhoptries, micronemes, dense granules, a conoid, apicoplast, nucleus, mitochondrion, ER, and Golgi. The subcellular architecture of *N. caninum* tachyzoites is critical to their infectious nature and their survival both outside and within host cells.

## Results

**Experimental layout of a 3D cryo-CDI microscope.** The CDI experiments were conducted at the refurbished ID10 beamline at the European Synchrotron Radiation Facility (ESRF) as summarized in Table 1. An X-ray beam with E = 8 keV was selected from the undulator radiation by a pseudo channel-cut Si(111) monochromator ($\Delta E/E = 1.4 \times 10^{-4}$). High order harmonics were suppressed by a pair of Si mirrors reflecting at the grazing incidence angle of 0.2°. A pair of roller blade slits closed to 7×7 µm² was placed 0.5 m upstream from the sample and produced coherent illumination at the sample. Diffraction from the beam defining slits was cleaned by a second pair of roller blade slits and finally by a Si corner placed in front of the sample. This setup produced a beam profile with a full width half maximum (FWHM) of ~9×9 µm² at the sample, with a beam flux of ~$1.46 \times 10^{10}$ photons/second. The coherent X-ray beam impinged on a silicon nitride membrane containing frozen-hydrated *N. caninum* cells, shown in Fig. 1. X-ray diffraction patterns were recorded using a MAXIPIX 2×2 detector[32], with a sample-to-detector distance of 5.29 m to meet the oversampling requirement[33,34]. The pixel size, detection area and image size of the MAXIPIX are 55×55 µm², 28.4×28.4 mm² and 516×516 pixels, respectively.

**3D data acquisition of frozen-hydrated cells.** Rapidly frozen *N. caninum* tachyzoites with cryo-protectants were supported on an X-ray transparent silicon nitride membrane and bathed in a 100K nitrogen gas stream to preserve the specimen during data acquisition (Fig. 1). Phase contrast optical microscopy images were used to confirm that rapidly frozen tachyzoites with cryo-protectants exhibited similar morphology to those parasites that were resuspended in a tissue culture medium (Fig. 2). The silicon nitride membrane was mounted onto a motorized stage with accurate 3D translation and 1D rotation capabilities, and scanned visually using an in-line optical microscope. Optical microscope images were correlated with diffraction patterns from the samples at all points scanned, in a mesh-like pattern. Cells identified on the membrane were chosen on the basis of their preliminary diffraction and visual inspection.

Once a cell was chosen, fine alignment was performed by diffraction – positioning the beam at the point in which the diffraction pattern was strongest from the sample. For this reason, it is important to prepare well-isolated cells with minimal background signal. Alignment by diffraction was performed for each angle, with an approximate exposure time of 5 seconds per angle to the sample. A single 2D diffraction pattern was collected for each angle using 8 keV X-rays with a 100 second exposure time. A total of 72 unique patterns were collected spanning approximately 111° (-60.6° to 50.9°, Supplementary Fig. 1), as a tomographic tilt series in agreement with the equally sloped tomography (EST) scheme[35]. 37 positive angle patterns and 34 negative were collected; at 0° the incident X-ray beam was approximately normal to the membrane, and data collection at high tilt angles was only limited by the geometry of the membrane support. The total acquisition time of the whole tilt series was approximately 5 hours. Improvements in the automated collection of diffraction patterns may reduce the time required for data collection to 2-3 hours per sample. A 0° pattern was collected multiple times during acquisition of the tomographic tilt series to assess diffraction quality of the sample and estimate radiation damage. A comparison of two independent 0° projections shows no appreciable difference (Supplementary Fig. 2). A rise in background intensity towards the end of our tomographic acquisition scheme reveals potentially high parasitic scattering from contamination or ice buildup near the sample (Supplementary Fig. 1). Otherwise, pattern features were generally conserved during data taking and the overall scattering intensity in patterns did not show appreciable decay (Supplementary Figs. 1 and 2).

**Post data processing and assembly of a 3D diffraction pattern.** To remove the background scattering, two diffraction patterns were acquired for every tilt angle by moving the cell in and out of the X-ray beam; again stressing the importance of a monodisperse, well isolated, preparation of cells with a clean surrounding background. The two diffraction patterns were scaled on a per-pixel basis, accounting for non-linearity and flat field correction to the detector[32]. The background pattern was then subtracted from the diffraction pattern of the cell to obtain corrected intensities. A portion of the low-resolution region of the resulting diffraction pattern, including the direct beam and the first speckle were blocked by a beam stop. Some of these pixels were recovered based on centro-symmetry, and all other pixels assigned the average of their symmetry counterparts. All 72 projections were processed in this way using a data analysis and quality control pipeline implemented in Matlab[33] (Fig. 3a and Supplementary Fig. 1). By combining the post-processed 2D diffraction patterns, a 3D matrix of diffraction intensities was assembled. The assignment of intensities within this matrix was achieved via an interpolation procedure using an inverse distance-weighting scheme.

**3D phase retrieval.** Phase retrieval of the assembled 3D diffraction pattern was achieved using the oversampling smoothness (OSS) algorithm[36]. OSS iterates back and forth between real and reciprocal space. In each iteration, a support is used as a constraint in real space and the measured data is enforced in reciprocal space. A cubic support was used in the initial reconstruction. The OSS algorithm exploits the correlation information among voxels in the region outside of the defined support in real space by applying a series of frequency filters in reciprocal space. In doing so, we stabilized our search for a global minimum in solution space. In addition, we broadened our search by conducting 100 independent reconstructions, each of which was monitored by $R_{rec}$

$$R_{rec} = \frac{\sum_{\vec{k}} \left\| F^e(\vec{k}) \right| - \gamma \left| F_j^c(\vec{k}) \right\|}{\sum_{\vec{k}} \left| F^e(\vec{k}) \right|} \qquad (1)$$

where $\left| F^e(\vec{k}) \right|$ is the experimental Fourier modulus (i.e. square root of the diffraction intensities), $\left| F_j^c(\vec{k}) \right|$ is the Fourier modulus calculated from a 3D reconstruction in the jth iteration, and $\gamma$ is a scaling factor. After 1,000 iterations, we obtained 100 independent 3D reconstructions, from which we selected the 10 best one with the lowest $R_{rec}$. By averaging these 10 reconstructions, we determined a tight 3D support, which is slightly larger than the cell boundary. Using this tight support, we ran another 100 independent reconstructions, each with 1,000 iterations. The final 3D structure was the average of the best 10 tight support reconstructions with the smallest $R_{rec}$ (approximately 25%).

**Estimation of 3D resolution achieved.** We quantified the resolution of our 3D reconstruction using two independent approaches. First, we calculated the phase retrieval transfer function (PRTF) of our 3D reconstructions[37] (Supplementary Fig. 3). Based on the PRTF=0.5 criterion, a resolution of approximately 68 nm was estimated. Second, we performed line scans along the three primary axes of the 3D reconstruction. A feature near the apical end of the cell provided strong contrast for estimating resolution. In the X and Y axes, a resolution of ~74 nm was measured (Fig. 4). Along the Z axis (*i.e.* the direction of the X-ray beam), which suffered from missing data, our resolution estimation was ~99 nm (Fig. 4). Collectively, these measurements indicate a resolution of ~75-100 nm (37-50 nm in the half period) was achieved in our 3D reconstruction.

**3D cryo-CDI reveals a *N. caninum* tachyzoite with a canonical polarized subcellular architecture.** The reconstructed 3D density shows a wedge-shaped cell with an overall size of ~1.5×2×4 μm (Fig. 3). Projections of the 3D density of the cell along its primary axes reveal characteristic views of a *N. caninum* tachyzoite. Figures 3b and c show a 2D projection and an isosurface rendering of the reconstructed 3D cell at the 0° tilt angle, respectively. These projection images resemble but exhibit a higher resolution than dark-field and bright-field optical microscope images of tachyzoites under similar conditions (Figs. 3d). Figure 3e shows a series of thin slices through the cell at a distance ranging from 250 to 1000 nm from the silicon nitride substrate, in which the red arrows indicate a conoid-like region at the apical end of the cell.

Definitive identification of cellular organelles was not possible from the recovered 3D density alone due in part to its limited resolution and the difficulty of segmenting nearly continuous features within its subcellular architecture. However, we constructed a 3D model from the tachyzoite density by outlining general features and boundaries for large subcellular structures, primarily focusing on the apical end of the cell (Fig. 5). Using semi-automatic segmentation, we distinguished between adjacent features based on their relative density, the size and appearance of the segmented objects (Methods). At the apical end of the cell, a bud-like structure was recognizable as the apical tip which has a tapered structure from which would emerge the conoid (Fig. 5). Underlying the apical tip, a network of electron-dense tubulovesicular structures are recognizable which likely correspond to the club-shaped rhoptries which are bundled together and tethered at the parasite's apex. Beneath this are two other internal structures that may represent the apicoplast and/or lobes of the single mitochondrion of the parasite. Finally, at the posterior end of the parasite, is a large oval structure corresponding to the parasite's nucleus. In total, five boundaries are assigned with characteristics described in detail in Table 2: including the cell membrane or boundary, an apical bundle resembling the rhoptries where tubules converge near the conoid[20], distinct but possibly interconnected networks of tubules that are more centrally located (mitochondria and/or apicoplast), and a nuclear region (Fig. 5 and Table 2). The volumes of these segmented regions range from 1.7% to 13.1% of the total cell volume (Table 2). The structures sit in close proximity to each other within the cell as can be appreciated from projected views of the traced out regions viewed from all three primary axes, and a series of projections at various angles along one of the primary axes (Fig. 5 and Supplementary video 1).

## Discussion

We have determined the 3D structure of a single cell in a frozen-hydrated state using cryo-CDI. The reconstructed 3D density of the isolated tachyzoite reveals its characteristic wedge shape with a 3D resolution of ~75 -100 nm. The distribution of the major organelles in the tachyzoite is polarized, with more electron dense structures packed near the apical end of the cell. From these structures we have identified the tachyzoite nuclear region, several dense networks of tubules resembling the rhoptries, and a conoid-like structure from which these tubules emanate at the apical tip of the cell.

The size and respective structures observed in the tachyzoite are in agreement with previously published negative-stain thin-section transmission electron microscopy (TEM) studies of *N. caninum* and *T. gondii* tachyzoites[38]. These TEM studies reported that micronemes, rhoptries, and clusters of dense granules are present in *N. caninum* cells[38].

Although we cannot definitively assign each of these structures in our 3D reconstruction, the sizes of structures we observed are consistent with those delineated from electron micrographs. The shape of our cell is distinctive of *N. caninum* tachyzoites, as observed in thin section TEM images[36]. The structure of the tachyzoite, as well as those observed by others[38,39], show a persistently unchanging outer membrane, whose bound is smooth and shows no ruffles. The most discernable subcellular compartments are a bundle of club-shaped rhoptries emanating from the conoid, a basket-shaped structure composed of microtubules at the apical end of the cell, and the nucleus, which sits mid-posterior within the parasite. Although the tachyzoite we observed is not perfectly round in its shortest dimension due to the geometric constraints imposed by the substrate, it retains its expected characteristic 3D shape.

It is important for future studies to consider improvements on both support geometry and methods for hours-long cryo-preservation of the sample. Since cumulative exposures are necessarily long for this type of experiment, contamination of background and/or sample are major concerns. Improvements on the geometry of sample supports aimed at minimizing the aqueous layer in which the sample is suspended and maintaining a constant, cold environment, offer marked improvements in background acquisition and the quality of diffraction from the sample. Keeping the cryo-CDI instrument in a helium chamber and using a liquid-nitrogen-cooled helium gas stream directed onto the sample will also reduce the ice build-up during the data acquisition. These improvements are sure to extend both the length of time allowed for tomographic data collection and the consistency of the data, resulting in more robust and high-resolution structures. In future studies, higher resolution would be a requisite step to better visualize organelles required for infection or the inner/outer membrane complexes of *N. caninum* or its apicomplexan relatives. This approach would thus be particularly well suited to the merozoite stage of the malarial parasite *Plasmodium falciparum*, which is smaller than both *N. caninum* and *T. gondii*[40].

In conclusion, this work demonstrates 3D imaging of a single frozen-hydrated *N. caninum* tachyzoite at a resolution of ~75-100 nm by cryo-CDI. We have achieved this without witnessing prohibitive radiation damage accumulated upon the sample in question and revealed the 3D morphology and subcellular architecture of this large and uniquely structured parasitic pathogen. This structure represents an experimental milestone toward high-resolution, lensless cryo-imaging of biological specimens using coherent X-rays. The resolution of this technique is ultimately limited by radiation damage to the samples. We now in 3D, and others previously in 2D[13,14], have shown that maintaining samples in a frozen-hydrated state can substantially mitigate radiation damage. Collectively, cryo-CDI efforts to date demonstrate a concerted effort toward realizing the imaging of whole cells at high resolution (~tens of nanometers) in three dimensions. Such an achievement is anticipated to help close the gap between the imaging of biological samples by electron and optical light microscopes. Lastly, as new, brighter, and highly coherent X-ray sources continue to emerge worldwide[42-45], our work presents a vision of what the field may ultimately achieve: the routine collection of high-resolution quantitative 3D structural information from cells in their native state. These studies are important building blocks for the better structural understanding of cells, and may lead to advances in the fields of imaging, biology, and medicine, as well as the general realm of biomaterials.

## Methods

**Preparation of *N. caninum* tachyzoites.** The NC1 strain *N. caninum* was grown and maintained by serial passage in confluent monolayers of human foreskin fibroblasts grown in DMEM supplemented with 10% fetal calf serum plus penicillin, streptomycin and glutamine[30]. A disperse suspension of *N. caninum* was prepared by extrusion from fibroblasts via passage through a syringe. A single cell suspension containing phosphate buffered saline, with 10% glycerol and 7% trehalose as cryo-protectants, was deposited on a 50-nm-thick silicon nitride membrane (Silson Ltd.), custom mounted on a brass pin support. Only those membranes with cells isolated on the membrane, visibly uncompromised and measured roughly 4-5 μm in their longest dimension were used for our imaging experiments.

**Rapid freezing of *N. caninum* cells.** Preparation of frozen-hydrated specimens requires rapid freezing of a thin aqueous layer to create vitreous ice. This is conventionally achieved by rapid 'plunge-freezing' into liquid ethane. This method has been widely used in cryo-electron microscopy (cryo-EM)[46]. Our attempts at plunge freezing of cells into liquid ethane and then transferring them into a nitrogen gas cryosteam showed variable results as the sample transferring step had a high probability of causing ice contamination. Ultimately, we found that high quality single cell diffraction was most reliably obtained when rapidly freezing cells directly into the nitrogen gas cryosteam. This method required optimization of the cryo-protectant cocktail, as described in the previous section. The cryo-protectants we chose have been shown to preserve the integrity of cells during freezing in liquid nitrogen and to stave off the nucleation and growth of ice crystals[47,48]. We further confirmed our protocol by comparing the morphology of *N. caninum* tachyzoites that were resuspended in a tissue culture medium, cryo-protectants only, and cryo-protectants following by rapidly freezing in liquid nitrogen. Our phase contrast microscopy images show that rapidly frozen tachyzoites with cryo-protectants exhibited similar morphology to parasites that were resuspended in a tissue culture medium and in cryo-protectants only (Fig. 2). However, there are drawbacks to the use of cryo-protectant agents, as these may thicken the vitreous ice layer and reduce contrast. Our future efforts will focus on uniform (high-pressure) freezing of samples, combined with focused ion beam milling, on substrates compatible with our instrument[36].

**Quantification of the radiation dose.** The X-ray flux was measured to be ~$1.46 \times 10^{10}$ photons/second within a $9 \times 9$ μm² illumination area at the sample. For a 100 second exposure, this equates to an X-ray flux of $P_t = 1.46 \times 10^{12}$ photons per projection on an area 81 μm², which corresponds to a flux per unit area $P_t / A = 1.8 \times 10^{10}$ photons/μm². The dose per projection, $D_p$, was estimated as

$$D_p = (P_t / A) \mu E / \rho, \qquad (2)$$

where $\mu$ is the linear absorption coefficient, $\rho$ the density of the cell, and $\mu/\rho$ was estimated to be 9.9 cm²/g with $E$ = 8 keV[28]. This gives a dose per projection $D_p$ of $2.28 \times 10^7$ Gy. The total dose ($D_t$) imparted onto the frozen-hydrated cell was estimated to be $1.64 \times 10^9$ Gy. With this dose, no noticeable deterioration was observed in the diffraction patterns (Supplementary Fig. 2). This is consistent with previous studies reporting no significant decay in the diffraction quality of samples after a dose of $5 \times 10^9$ Gy in a frozen-hydrated state[13,14,28,29].

**3D model building and rendering.** The final 3D reconstruction was processed to obtain a 3D model representing the cell bounds and the regions within the cell that may pertain to specific subcellular structures or organelles. The density per voxel of the 3D reconstruction was used as a metric for demarcating regions. The IMOD suite was employed for this and subsequent model building tasks[49]. The program includes the 3dmod package, and facilitates semi-automatic model building, display, and image processing on a per-slice basis, to build a 3D model from the 3D cell reconstruction[49]. This approach has been successfully applied to electron microscopy images[50]. We relied on several manually chosen thresholds for semi-automatic segmentation. The cell bound was chosen using a threshold that excluded all voxels with density less than approximately 10% that of the maximum density in the reconstruction. For regions within the cell, local contrast was used to outline objects of interest and their bounds. For each object, sections in between the manually outlined closed contours were interpolated automatically using 3dmod. The final images were rendered using the image processing programs ImageJ, Amira and Chimera, and assembled into figures using Adobe Illustrator.

## References

1. Stephens, D. J. & Allan, V. J. Light Microscopy Techniques for Live Cell Imaging. *Science* **300**, 82-86 (2003).

2. Huang, B., Bates, M. & Zhuang, X. Super resolution fluorescence microscopy. *Annu. Rev. Biochem.* **78**, 993-1016 (2009).

3. Scott, M. C. *et al*. Electron tomography at 2.4-ångström resolution. *Nature* **483**, 444–447 (2012).

4. Chen, C. C. *et al*. Three-dimensional imaging of dislocations in a nanoparticle at atomic resolution. *Nature* **496**, 74-77 (2013).

5.  Lučić, V., Förster, F., Baumeister, W. Structural Studies by Electron Tomography: From Cells to Molecules. *Annu. Rev. Biochem.* **74**, 833-865 (2005).

6.  Miao, J., Charalambous, P, Kirz, J. & Sayre, D. Extending the methodology of X-ray crystallography to allow imaging of micrometre-sized non-crystalline specimens. *Nature* **400**, 342-344 (1999).

7.  Chapman, H.N. & Nugent, K.A. Coherent lensless X-ray imaging. *Nat. Photon.* **4**, 833–839 (2010).

8.  Miao, J. *et al.* Imaging whole Escherichia coli bacteria by using single-particle x-ray diffraction. *Proc. Natl Acad. Sci. USA*. **100**, 110-112 (2003).

9.  Shapiro, D. *et al.* Biological imaging by soft x-ray diffraction microscopy. *Proc. Natl Acad. Sci. USA*. **102**, 15343-15346, (2005).

10. Jiang, H. *et al.* Nanoscale Imaging of Mineral Crystals inside Biological Composite Materials Using X-Ray Diffraction Microscopy. *Phys. Rev. Lett.* **100**, 038103 (2008).

11. Song, C. *et al.* Quantitative Imaging of Single, Unstained Viruses with Coherent X Rays. *Phys. Rev. Lett.* **101**, 158101 (2008).

12. Nishino, Y. *et al.* Three-Dimensional Visualization of a Human Chromosome Using Coherent X-Ray Diffraction. *Phys. Rev. Lett.* **102**, 018101 (2009).

13. Huang, X. *et al.* Soft X-Ray Diffraction Microscopy of a Frozen Hydrated Yeast Cell. *Phys. Rev. Lett.* **103**, 198101 (2009).

14. Lima, E. *et al.* Cryogenic X-Ray Diffraction Microscopy for Biological Samples. *Phys. Rev. Lett.* **103**, 198102 (2009).

15. de la Cuesta, F. B. *et al*. Coherent X-ray diffraction from collagenous soft tissues. *Proc. Natl Acad. Sci. USA*. **106**, 15297–15301 (2009).

16. Jiang, H. *et al.* Quantitative 3D imaging of whole, unstained cells by using X-ray diffraction microscopy. *Proc. Natl Acad. Sci. USA*. **107**, 11234-11239 (2010).

17. Nelson, J. *et al.* High-resolution x-ray diffraction microscopy of specifically labeled yeast cells. *Proc. Natl Acad. Sci. USA.* **107**, 7235-7239 (2010).

18. Giewekemeyer, K. *et al.* Quantitative biological imaging by ptychographic x-ray diffraction microscopy. *Proc. Natl Acad. Sci. USA.* **107**, 529-534, (2010).

19. Daewoong, N. *et al.* Imaging Fully Hydrated Whole Cells by Coherent X-Ray Diffraction Microscopy. *Phys. Rev. Lett.* **110**, 098103 (2013).

20. Kimura, T. *et al.* Imaging live cell in micro-liquid enclosure by X-ray laser diffraction. *Nat. Commun.* **5**, 3052 (2014).

21. Gallagher-Jones, M. *et al.*, Macromolecular structures probed by combining single-shot free-electron laser diffraction with synchrotron coherent X-ray imaging. *Nat. Commun.* **5**, 3798 (2014).

22. Bergh, M., Huldt, G., Tîmneanu, N., Maia, F. R. N. C. & Hajdu, J. Feasibility of imaging living cells at sub-nanometer resolution by ultrafast X-ray diffraction. *Q. Rev. Biophys.* **41**, 181–204 (2008).

23. Seibert, M. M. *et al.* Single mimivirus particles intercepted and imaged with an X-ray laser. *Nature* **470**, 78–81 (2011).

24. Schlichting, I. & Miao, J. Emerging opportunities in structural biology with X-ray free-electron lasers. *Curr. Opin. Struct. Biol.* **22**, 613-26 (2012).

25. Kirz, J., Jacobsen, C. & Howells, M. Soft X-ray microscopes and their biological applications. *Q. Rev. Biophys.* **28**, 33-33 (1995).

26. Le Gros, M. A., McDermott, G., Larabell, C. A. X-ray tomography of whole cells. *Curr. Opin. Struct. Biol.* **15**, 593–600 (2005).

27. Meirer, F., Cabana, J., Liu, Y., Mehta, A., Andrews, J. C., & Pianetta, P. Three-dimensional imaging of chemical phase transformations at the nanoscale with full-field transmission X-ray microscopy. *J. Synchrotron Radiat.* **18**, 773–781 (2011).

28. Howells, M. R. *et al*. An assessment of the resolution limitation due to radiation-damage in X-ray diffraction microscopy. *J. of Electr. Spectr. Related Phenom*. **170**, 4–12 (2009).

29. Shen, Q. *et al*. Diffractive imaging of nonperiodic materials with future coherent X-ray sources. *J. Synchrotron Radiat.* **11**, 432-438 (2004).

30. Sohn, C. S. *et al*. Identification of novel proteins in *Neospora caninum* using an organelle purification and monoclonal antibody approach. *PLoS One*. **6**, e18383 (2011).

31. Carruthers, V.B. & Suzuki, Y. Effects of *Toxoplasma gondii* infection on the brain. *Schizophr. Bull*. **33**, 745-751 (2007).

32. Ponchut, C. *et al*. MAXIPIX, a fast readout photon-counting X-ray area detector for synchrotron applications. *J. Instrument*. **6**, C01069 (2011).

33. Miao, J., Sayre, D. & Chapman, H. N. Phase Retrieval from the Magnitude of the Fourier transform of Non-periodic Objects. *J. Opt. Soc. Am. A* **15**, 1662-1669 (1998).

34. Miao, J., Ishikawa, T., Anderson, E. H. & Hodgson, K. O. Phase retrieval of diffraction patterns from noncrystalline samples using the oversampling method. *Phys. Rev. B* **67**, 174104 (2003).

35. Miao, J., Foster, F., Levi, O., 2005. Equally Sloped Tomography with Oversampling Reconstruction. *Phys. Rev. B* **72**, 052103.

36. Xu, R. *et al*. Coherent diffraction microscopy at SPring-8: instrumentation, data acquisition and data analysis. *J. Synch. Rad*. **18**, 293-298 (2011).

37. Rodriguez, J. A. *et al*. Oversampling smoothness: an effective algorithm for phase retrieval of noisy diffraction intensities. *J. Appl. Crystallogr*. **46**, 312-318 (2013).

38. Chapman, H. N. *et al*. High-resolution *ab initio* three-dimensional x-ray diffraction microscopy. *J. Opt. Soc. Am. A* **23**, 1179-1200 (2006).

39. Speer, C. A. *et al*. Comparative ultrastructure of tachyzoites, bradyzoites, and tissue cysts of *Neospora caninum* and *Toxoplasma gondii*. *Int. J. Parasitol*. **29**, 1509-1519 (1999).

40. Schatten, H. & Ris H. Unconventional Specimen Preparation Techniques Using High Resolution Low Voltage Field Emission Scanning Electron Microscopy to Study Cell Motility, Host Cell Invasion, and Internal Cell Structures in *Toxoplasma gondii*. *Microscopy and Microanalysis*. **8**, 94-103 (2002).

41. Baum, J., Gilberger, T. W., Frischknecht, F. & Meissner, M. Host-cell invasion by malaria parasites: insights from Plasmodium and Toxoplasma. *Trends Parasitol.* **24**, 557-63 (2008).

42. Emma, P. *et al*. First lasing and operation of an ångstrom-wavelength free-electron laser. *Nat. Photon*. **4**, 641–647 (2010).

43. Ishikawa, T. *et al.* A compact X-ray free-electron laser emitting in the sub-ångström region. *Nat. Photon*. **6**, 540–544 (2012).

44. Popmintchev, T. *et al.* Bright Coherent Ultrahigh Harmonics in the keV X-ray Regime from Mid-Infrared Femtosecond Lasers. *Science* **336**, 1287–1291 (2012).

45. BESAC Subcommittee on Future Light Sources. Grand Challenge Science on Diffraction Limited Storage Rings. Available at www.aps.anl.gov/Upgrade/Documents/DLSR_report_for_BESAC.pdf (2013).

46. Dubochet, J. *et al*. Cryo-electron microscopy of vitrified specimens. *Q. Rev. Biophys.* **21**, 129-228 (1998).

47. Pellerin-Mendez, C. *et al.* In Vitro Study of the Protective Effect of Trehalose and Dextran during Freezing of Human Red Blood Cells in Liquid Nitrogen. *Cryobiology*. **35**, 173-186 (1997).

48. Rowe, A. W. *et al*. Liquid nitrogen preservation of red blood cells for transfusion. *Cryobiology*. **5**, 119-128 (1968).

49. Kremer, J. R. et al. Computer Visualization of Three-Dimensional Image Data Using IMOD. *J. Struct. Biol.* **116**, 71-76 (1996).

50. Nicastro, D. *et al*. The Molecular Architecture of Axonemes Revealed by Cryoelectron Tomography. *Science* **313**, 944-948 (2006).

**Acknowledgements**

This work is supported by the DARPA PULSE program through a grant from AMRDEC and the National Institutes of Health (grant # GM081409-01A1). P.J.B. thanks the supported by the National Institutes of Health (R01#AI064616). J.A.R. acknowledges the support of the Howard Hughes Medical Institute Gilliam Fellowship for graduate studies, the UCLA MBI Whitcome Fellowship, and the A.P. Giannini Postdoctoral fellowship. H. J. acknowledges the support of the National Natural Science Foundation of China (21390414).

**Figures Legends**

**Figure 1. Schematic layout of a 3D cryo-CDI microscope.** (**a**) A diagram shows a silicon-nitride-membrane containing frozen-hydrated cells on a single-tilt piezo-electric stage, bathed in a nitrogen gas cryosteam. A coherent X-ray beam impinges on a cell, from which a tilt series of 72 diffraction patterns was measured using a MAXIPIX detector (**b**). A 3D diffraction pattern was assembled from the 72 2D diffraction patterns, from which the 3D structure of the cell (**c**) was iteratively reconstructed by using the oversampling smoothness (OSS) algorithm[37].

**Figure 2**. **Examination of the cell morphology of rapidly frozen *N. caninum* with cryo-protectants.** Extracellular *N. caninum* tachyzoites were resuspended in a tissue culture medium (**a**), cryo-protectants only (**b**), and cryo-protectants following by rapid freezing in liquid nitrogen (**c**). The cells were imaged with an AxioCam MRm CCD camera and AxioVision software on an Axio Imager Z1 microscope (Zeiss) using a 100x oil immersion objective. Rapidly frozen tachyzoites with cryo-protectants exhibited similar morphology to those parasites that were resuspended in cryo-protectants only and in a tissue culture medium. Scale bar: 5 µm.

**Figure 3. 3D structure determination of a whole, frozen-hydrated *N. caninum* cell.** (**a**) A representative diffraction pattern taken at the 0° tilt angle in which the horizontal and vertical bars are caused by the tiling of the 2x2 modules of the MAXIPIX detector[30]. An inset shows an enlarged version of the low spatial frequency region of the pattern where the mising data at the center is due to a beam stop. A 2D projection (**b**) and an isosurface rendering (**c**) of the recosntructed 3D cell at the 0° tilt angle. (**d**) Dark-field and bright-field optical microscope images of similar cells are shown enlarged to an equivalent scale for comparison. (**e**) A series of thin slices through the cell at a distance ranging from 250 to 1000 nm away from the silicon nitride substrate, in which the red arrows indicate a conoid-like region. Images in (**a, b, e**) are false colored: red, yellow, green, blue and dark blue/black range from high, medium and low to no signal/density. Scale bars: 500 nm.

**Figure 4. Resolution estimation for the 3D cryo-CDI reconstruction.** (**a, b**) Two thin slices through the 3D reconstruction of the cell. Dotted lines c*, d* and e* correspond to line scans along the X, Y and Z axes, respectively, where the X-ray beam is in the Z axis. Images are false colored; red, yellow, green, blue and dark blue/black range from high, medium and low to no signal/density, repsectively. Scale bar: 500 nm. (**c-e**) Line scans of c*, d* and e* indicate a resolution of ~74 nm, ~74 nm, and ~99 nm was achieved in the 3D reconstruction along the X, Y and Z axes, respectively. The lower resolution along the Z axis is due to the missing data (wedge) in that direction.

**Figure 5. 3D architecture of the frozen-hydrated *N. caninum* cell.** (**a**) A deconstructed isosurface model of the *N. caninum* tachyzoite is shown with 2D projections along the beam direction of 3D renderings of each region chosen to be highlighted (see Table 2). At the center, the boundary of the cell is shown, as well as each of the regions in its corresponding location within the cell. Regions are numbered and colored according to their listing in Table 2. A line extends from each of these regions to their cutouts, which are numbered and bound by the three lengths of the box that circumscribes them; scale bars: 500 nm. (**b**) Views of the 3D isosurface model of the tachyzoite along the three primary axes. The Z-axis corresponds to a view along the beam direction at the 0° tilt angle. (**c**) Starting from the 0° tilt angle, the Z-axis view in (**b**) is rotated about a primary axis, each frame (1-6) in the montage is a 72° tilt apart from the previous.

## Tables

**Experiment Design and Data Analysis**

**Sample and Environment**

| | |
|---|---|
| Sample (strain) | *N. caninum* tachyzoite (NC1) |
| Host cell | Human foreskin fibroblasts |
| Preparation method | Centrifugation after normal egress |

*Ambiance*

| | |
|---|---|
| Support | $Si_3N_4$ membrane (50 nm thick) |
| Temperature | 100 K (cryostream) |
| Local Environment | Thin liquid film |
| Buffer composition | PBS, 10% Glycerol, 7% Trehalose |

**Instrumentation and Experiment Design**

| | | ID10C Beamline, ESRF |
|---|---|---|
| Workstation | | |
| *Beam Parameters* | | |
|    Energy (wavelength) | | 8 KeV (1.55 Å) |
|    Beam slit size | | 7×7 μm$^2$ |
|    Distance to sample | | 0.5 m |
|    FWHM at sample | | 9×9 μm$^2$ |
|    Flux at sample | | 1.46×10$^{10}$ photons/second |

**Data Collection**

| | |
|---|---|
| Number of projections | 72 |
| Exposure time per projection | 100 seconds |
| Sample to detector distance | 5.29 m |
| Linear oversampling ratio | $O_x$=3.3, $O_y$=6.0, $O_z$=8.1 |
| Tilt range | -60.6° – +50.9° |
| Total radiation dose | 1.64×10$^9$ Gy |

**Phasing Method and Structure Rendering**

| | |
|---|---|
| Primary processing and analysis | Custom MatLab script |
| Phase retrieval algorithm (data type) | OSS (assembled 3D matrix) |
| Number of Iterations (independent seeds) | 1,000 (100) |
| Segmentation (rendering software) | Manual, threshold-based (Amira, 3Dmod, Chimera) |

**Table 1**. Summary of sample and experiment parameters, data statistics, and results.

| | Region | Color | Area (μm$^2$) | Volume (μm$^3$) | % of Cell Volume |
|---|---|---|---|---|---|
| 1 | Cell Boundary | Pearl | 20.4 | 5.7 | 100 |
| 2 | Rhoptries | Red | 4.8 | 0.4 | 6.5 |
| 3 | Apicoplast | Orange | 3.1 | 0.3 | 4.9 |
| 4 | Nucleus | Brown | 8.6 | 0.8 | 13.1 |
| 5 | Mitochondrion | Blue | 1.9 | 0.1 | 1.7 |

**Table 2.** Summary of the five regions comprising the isosurface model of the reconstructed 3D tachyzoite. Regions are numbered according to their presentation in Fig. 5.

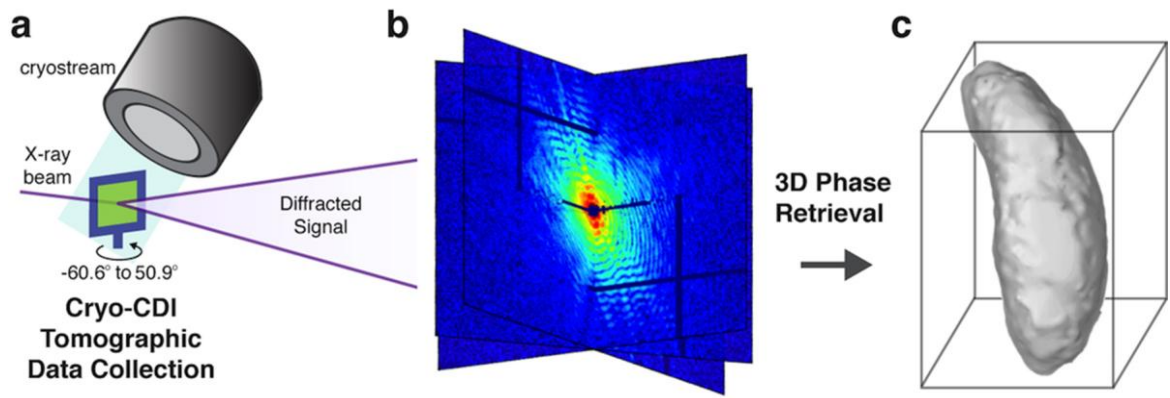

FIG. 1

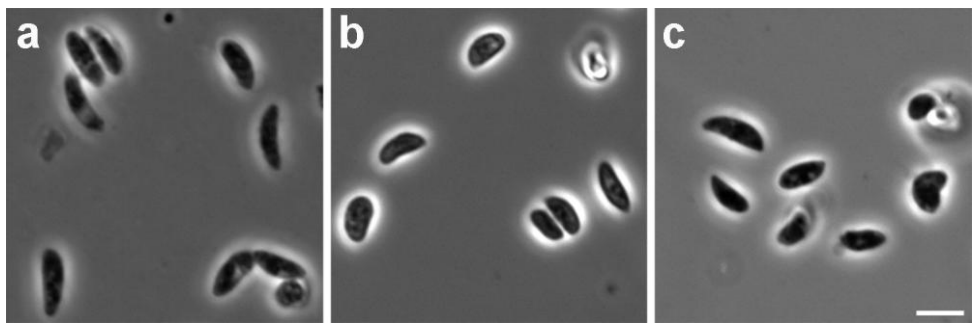

FIG. 2

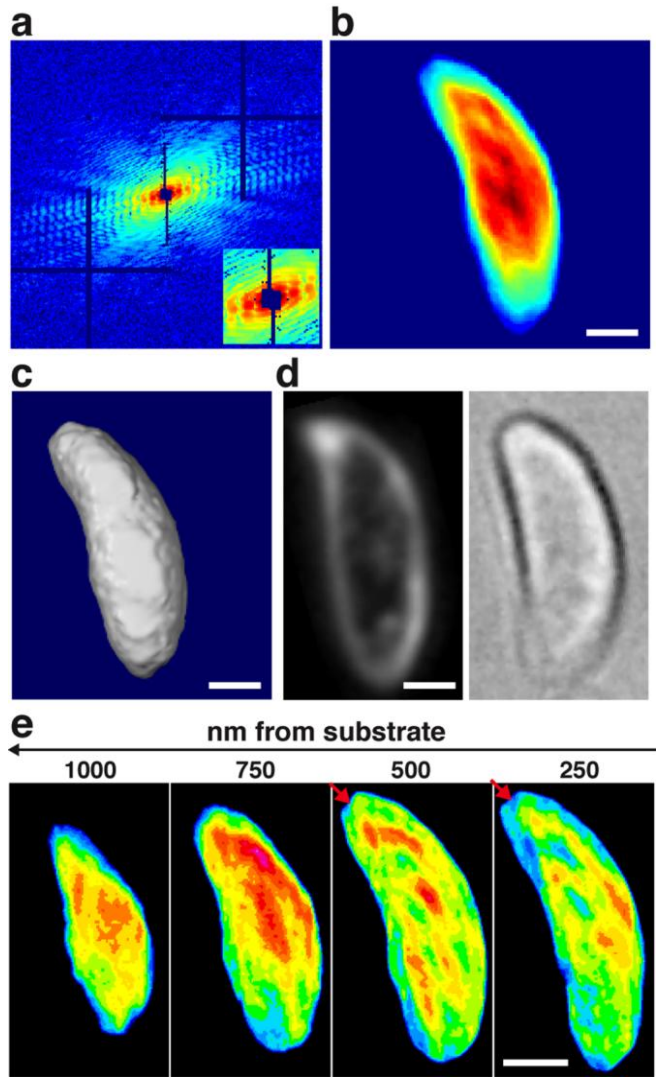

FIG. 3

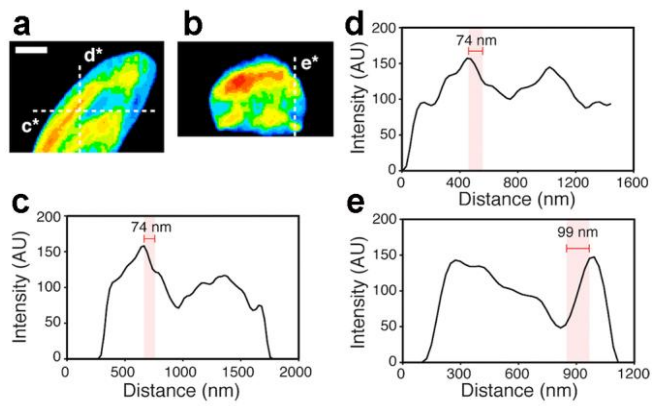

FIG. 4

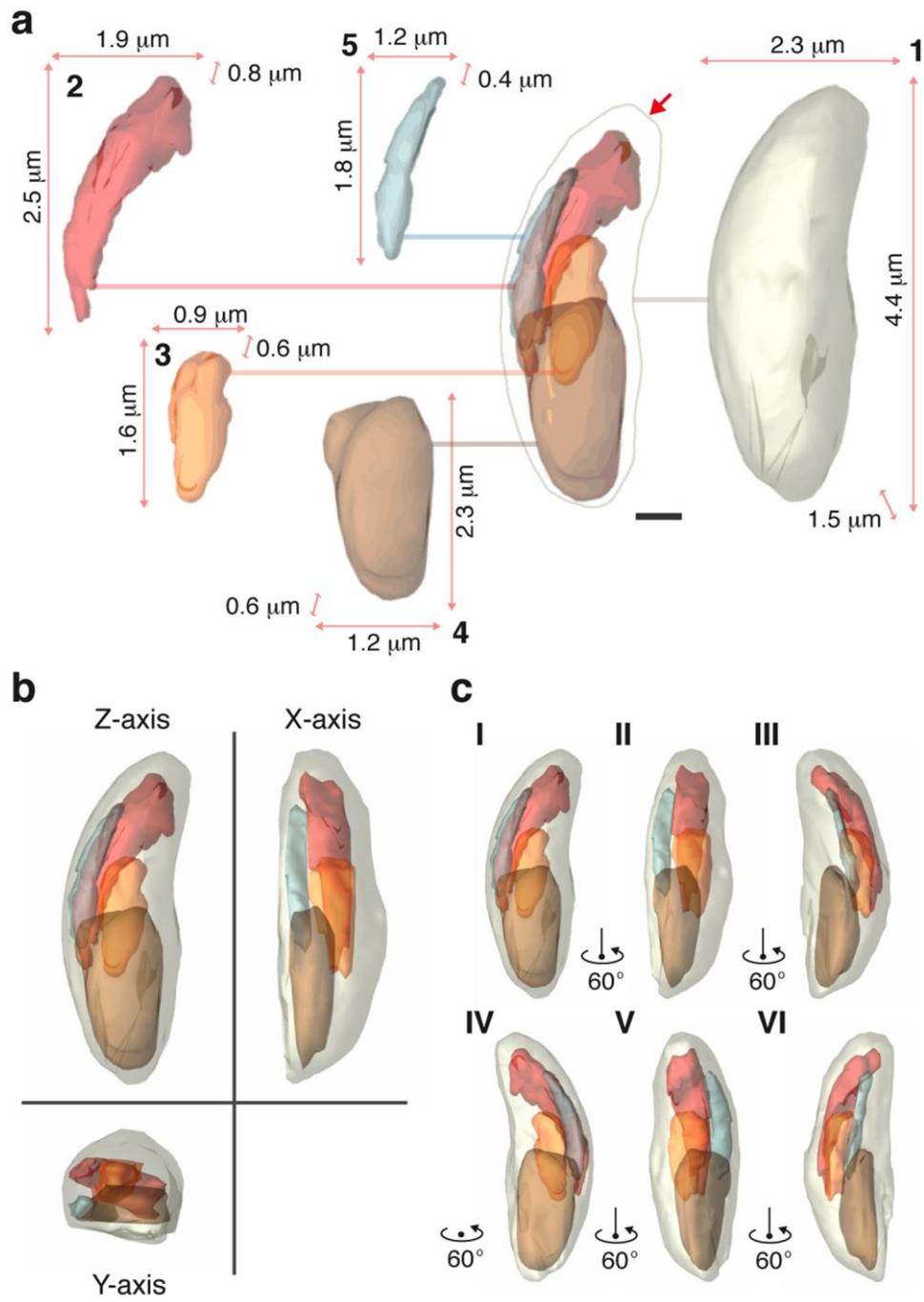

FIG. 5